\documentclass[epj,final]{svjour}
\usepackage{psfig}

\begin{document}
\title{Transport analysis of K$^+$ production in proton-nucleus reactions}
\author{Z.
Rudy\inst{1}            \and W. Cassing\inst{2}   \and L.
Jarczyk\inst{1}         \and B. Kamys\inst{1}   \and A.
Kowalczyk\inst{1,2} \and P. Kulessa\inst{3}
}                     
\institute{ M. Smoluchowski Institute of Physics, Jagellonian
University, PL-30059 Cracow, Poland \and
Institut f\"ur Theoretische Physik, Justus Liebig Universit\"at
Giessen, D-35392 Giessen, Germany    \and
W. Niewodnicza\'nski Institute of Nuclear Physics, PL-31342, Cracow,
Poland }

\date{Received: date / Revised version: date}
\abstract{ The production of $K^+$  mesons in proton-nucleus
collisions from 1.0 to 2.3 GeV  is analyzed with respect to
one-step nucleon-nucleon
 $(NN\rightarrow N Y K^+$)
 and two-step $\Delta$-nucleon $(\Delta N \rightarrow  K^+ Y N$) or
 pion-nucleon  $(\pi N \rightarrow  K^+ Y $)
 production channels on the basis of a coupled-channel transport
 approach (CBUU) including the kaon final-state-interactions (FSI).
Momentum-dependent potentials for the nucleon,
hyperon and kaon in the final state are included as well as
 $K^+$ elastic rescattering in the target nucleus.
The transport calculations are compared to the experimental $K^+$
spectra taken at COSY-J\"ulich. Our systematic analysis of $K^+$
spectra from $^{12}C$, $^{63}Cu$, $^{107}Ag$  and $^{197}Au$ targets as well as
their momentum differential ratios gives a repulsive $K^+$
potential of $20\pm 5 $ MeV at normal nuclear matter density.
\PACS{
      {13.60.Le}{Meson production}   \and
      {13.75.Jz}{Kaon-baryon interactions} \and
      {14.40.Aq}{Pi, K, and eta mesons}    \and
      {24.40.-h}{Nucleon-induced reactions}
     } 
} 

\maketitle

\section{Introduction}
The production of heavy mesons in $p+A$ reactions at
bombarding energies far below and close to the free
nucleon-nucleon threshold is of specific interest
\cite{1}-\cite{cass95} as one hopes to learn either about
cooperative nuclear phenomena and/
or about high momentum
components of the nuclear many-body wave function that arise from
nucleon-nucleon correlations. Especially $K^+$ mesons have been
considered as promising hadronic probes \cite{5,12,Geiss} due to the
rather moderate final state interaction, which is a consequence of
 strangeness conservation.

Experiments on $K^\pm$ production from nucleus-nucleus collisions
at SIS energies of 1-- 2 A$\cdot$ GeV have shown that in-medium
properties of the kaons are seen in the collective flow pattern of
$K^+$ mesons both, in-plane and out-of-plane, as well as in the
abundancy  of antikaons \cite{cass99,Li2001,SengerPPNP}. Thus in-medium
modifications of the mesons have become a topic of substantial
interest in the last decade triggered in part by the early
suggestion of Brown and Rho \cite{BR}, that the modifications of
hadron masses should scale with the scalar quark condensate
$<q\bar{q}>$ at finite baryon density.

As demonstrated in the pioneering work of Kaplan and Nelson
\cite{nelson,Kaplan} kaons and antikaons couple attractively to
the scalar nucleon density with a strength proportional to the
$KN- \Sigma$ constant $\Sigma_{KN}$, which is not well known at
present and may vary from 270 to 450 MeV. Furthermore, a vector
coupling to the quark 4-current -- for vanishing spatial
components -- leads to a repulsive potential term for the kaons
and to an additional attractive term for the antikaons. In fact,
the recent observation of a rather narrow strange tribaryon state
with mass 3.115 GeV in Ref. \cite{suzuki} -- observed in the
$^4He(K^-, p)X$ reaction with antikaons at rest -- has
given further experimental evidence for strong attractive $K^-$
potentials in (light) nuclei.

Nevertheless, the actual kaon and antikaon self energies (or
potentials) at normal nuclear matter density (and above)
are quite a matter of debate -- due to higher order
terms in the chiral expansion -- especially for the antikaon
\cite{Gal,Lutz,Ramos}. Moreover the momentum-dependence of their self
energies  is widely unknown (except for a dispersion analysis in
Ref. \cite{Sib98}) since most Lagrangian models are restricted to
$s$-wave interactions or only include additional $p$-waves. It is
thus mandatory to perform experimental studies of the (anti-) kaon
properties under well controlled conditions, e.g. in
proton-nucleus reactions, where one probes the (anti-) kaon self
energies  at normal nuclear matter density $\rho_0
\approx$ 0.16 fm$^{-3}$ and below. Furthermore, by gating on kaon momenta in
the laboratory, one might be able to obtain information on the
momentum dependence of the self energies, too.

 $K^+$ production in $p+A$ collisions at subthreshold energies
has been observed experimentally more than a decade ago by Koptev
et al. \cite{12} at bombarding energies from 0.8 to 1.0 GeV.
However, only total $K^+$ yields could be extracted at that time
at subthreshold energies.
In more recent years differential $K^+$ spectra
have been measured down to 1.2 GeV for $^{12}C$ targets at 40 $^0$
\cite{debowski} (SATURNE) or 90 $^0$ in the laboratory
\cite{badala} (CELSIUS). Unfortunately, the different
experiments have no overlap in acceptance and the interpretation
of the data, if compatible at all, remains vague \cite{Markus,zibi02}.
First data on the full momentum distribution at forward angles
have been presented  by the ANKE Collaboration at
COSY-J\"ulich \cite{Barsov} for $K^+$ mesons from $p+^{12}C$
reactions at 1.0 GeV in 2001 \cite{ANKE} and at higher bombarding energies
in the last year \cite{Markus04}.

In this study we use the coupled-channel (CBUU) transport model that
has been developed first in Ref. \cite{Wolf} for the description of
nucleus-nucleus collisions and later on employed for the simulation
of pion- and proton-nucleus reactions \cite{Sib98,zibi95,cass98},
too. For applications to $K^\pm$ production in nucleus-nucleus
collisions at SIS energies we refer the reader to Refs.
\cite{cass97}. In this model the effects of momentum-dependent self
energies for all hadrons can be studied explicitly as well as their
production and propagation in the nuclear medium. The actual version
of the CBUU transport model employed in our present analysis is
identical to that described in Ref. \cite{zibi02}. We thus discard
an explicit description and refer the reader to Ref. \cite{zibi02}
for technical details and the explicit momentum-dependent potentials
employed for the nucleon and hyperon degrees of freedom.

The paper is organized as follows: We very briefly recapitulate
the concepts of the CBUU model in Section 2 and step on with the
actual results for $K^+$ spectra in $p+A$ collisions at COSY
energies in Section 3 in comparison to the data of the ANKE
collaboration \cite{Markus04}. In  Section 4 we compare the
calculated ratios of $K^+$ spectra for different kaon potentials
with the corresponding data at various bombarding energies
\cite{Markus04}. This systematic ana\-lysis will lead to an
approximate extraction of the strength of the $K^+$ potential in
the nuclear medium. A summary
 concludes this paper in Section 5.

\section{Brief reminder of the CBUU  transport model}

In this work we perform our analysis along the line of the
CBUU\footnote{Coupled-Channel Boltzmann-Uehling-Uhlenbeck}
 approach~\cite{Wolf} which is based on a
coupled set of  transport equations for the phase-space
distributions $f_{h} (x,p)$ of hadron $h$, i.e.
\cite{weber}
\begin{eqnarray}
&& \left( \Pi_{\mu}-\Pi_{\nu}\partial_{\mu}^p
U_{h}^{\nu} -M_{h}^*\partial^p_{\mu} U_{h}^{S}
\right) \partial_x^{\mu} f_{h}(x,p) \label{g24} \\
&&+ \left( \Pi_{\nu} \partial^x_{\mu}
U^{\nu}_{h}+ M^*_{h} \partial^x_{\mu}U^{S}_{h}\right)
\partial^{\mu}_p  f_{h}(x,p)   \nonumber\\
&&= \sum_{h_2 h_3 h_4\ldots} \int d2 d3 d4
 [G^{\dagger}G]_{12\to 34\ldots} \nonumber \\
&&\phantom{a}\hspace*{2cm}\times\delta^4(\Pi +\Pi_2-\Pi_3-\Pi_4  ) \nonumber\\
&& \times \left\{ f_{h_3}(x,p_3)f_{h_4}(x,p_4)\bar{f}_{h}(x,p)
\bar{f}_{h_2}(x,p_2) \right.\nonumber \\
&& - \left. f_{h}(x,p)f_{h_2}(x,p_2)\bar{f}_{h_3}(x,p_3) \bar{f}_{h_4}(x,p_4)
\right\}  \ .
\end{eqnarray}
In Eq.~(\ref{g24}) $U_{h}^{S}(x,p)$ and $U_{h}^{\mu}(x,p)$ denote
the real part of the scalar and vector hadron selfenergies,
respectively, while $[G^\dagger G]_{12\to 34\ldots} \delta^4 (\Pi
+\Pi_2-\Pi_3-\Pi_4\ldots )$ is the 'transition rate' for the
process $1+2\to 3+4+\ldots$ which is taken to be on-shell in the
semiclassical limit adopted. The hadron quasi-particle properties
in (\ref{g24}) are defined via the mass-shell constraint
\begin{equation}   \label{g25}
\delta (\Pi_{\mu}\Pi^{\mu}-M_{h}^{*2} ) \ \ ,
\end{equation}
with effective masses and momenta (in local Thomas-Fermi
approximation) given by \cite{weber}
\begin{eqnarray}\label{g26}
M_{h}^* (x,p)&=&M_h + U_h^{{S}}(x,p) \nonumber \\ \Pi^{\mu}
(x,p)&=&p^{\mu}-U^{\mu}_h (x,p)\ \ ,
\end{eqnarray}
while the phase-space factors
\begin{equation}
\bar{f}_{h} (x,p)=1 \pm f_{{h}} (x,p)
\end{equation}
are responsible for fermion Pauli-blocking or Bose enhancement,
respectively, depending on the type of hadron in the final/initial
channel.  The transport approach
(\ref{g24}) is fully specified by $U_{h}^{S}(x,p)$ and
$U_{h}^{\mu}(x,p)$ $(\mu =0,1,2,3)$, which determine the
mean-field propagation of the hadrons, and by the transition rates
$G^\dagger G\,\delta^4 (\ldots )$ in the collision term (5), that
describes the scattering and hadron production/\-absorption rates.

The scalar and vector mean fields $U_{h}^{S}$ and $U^\mu_{h}$ for
nucleons are modeled  in line with Ref. \cite{excita} and
presented explicitly in Fig. 1 of Ref. \cite{zibi02} for the
resulting real part of the nucleon potential $U_N(p)$. We recall
that this nucleon potential (at density $\rho_0$) is attractive
for momenta $p$ below 0.6 GeV/c and repulsive above. Apart from
the nuclear potentials each charged hadron additionally moves in
the background of the Coulomb potential that is generated by the
charged hadrons themselves. In case of proton-nucleus reactions --
with the nucleus at rest -- this essentially amounts to a Coulomb
acceleration in the final state for positively charged hadrons and
a deceleration for negatively charged particles. Note, that for
heavy nuclei like $Pb$ or $Au$ the Coulomb potential in the
nuclear interior (for $p, \pi^+, K^+$) is $\sim$ 20 MeV, i.e. of
the same order of magnitude as the 'expected' repulsive $K^+$
nuclear potential \cite{zibi02,Mischa}.

 The hyperon mean fields, furthermore, are assumed to be 2/3
of the nucleon potentials. In the present approach, apart from
nucleons, $\Delta, N(1440)$, $N(1535)$, $\Lambda, \Sigma$ with
their isospin degrees of freedom, we propagate explicitly pions,
$K^+$, and $\eta$'s and assume that the pions as Goldstone bosons
do not change their properties in the medium.  The kaon potential
is taken as momentum independent (cf. Ref. \cite{zibi02}), while
its strength is varied in the dynamical calculations from 0 to 20
MeV at nuclear matter density $\rho_0$ assuming a linear increase
with the actual nucleon density $\rho({\bf r})$.

The calculation of 'subthreshold' particle production is
treated perturbatively in the energy regime of interest due to the
small cross sections involved. Since we work within the parallel
ensemble algorithm in the CBUU approach, each parallel run of the
transport calculation
can be considered approximately as an individual reaction event,
where binary reactions in the entrance channel at given invariant
energy $\sqrt{s}$ lead to final states with 2 (e.g. $K^+ Y$ in
$\pi B$ channels), 3 (e.g. for $K^+ YN$ channels in $BB$
collisions) or 4 particles (e.g. $K\bar{K}NN$ in $BB$ collisions)
with a relative weight $P_i$ for each event $i$ which is defined
by the ratio of the  production cross section to the total
hadron-hadron cross section. We thus
dynamically gate on all events where a $K^+$  meson
is produced initially.  Each strange hadron production event $i$
is represented by  $\sim 10^3$ testparticles for the final strange
hadron $j$ with individual weight $W_j^i$ such that the sum of the
weights $W_j^i$ over $j$ reproduces the individual production
probability $P_i$ and the distribution in momenta (multiplied with
the $NN$ or $\pi N$ cross section) describes the differential
production cross section (cf. Ref. \cite{zibi02} for details).

For the present study the production of pions by $pN$ collisions
in $p+A$ reactions as well as the total kaon cross sections in $pN$
and $\pi N$ collisions are of relevance. The pion production cross
section from $NN$ interactions is based on the parametrization of
the experimental data by Ver West and Arndt \cite{21} and
implemented in the transport model as described in Ref.
\cite{Wolf}. The cross sections for the channels $\pi N
\rightarrow K Y$, where $Y$ stands for a hyperon $\Lambda,
\Sigma$, are taken from the analysis of Huang et al. \cite{huang}
and essentially correspond to the experimental data for the
different $\pi N$ channels in vacuum (or 'free' space). Note, that
in addition to the early studies in \cite{15,cass95} the channels
with a $\Sigma$ hyperon are  taken into account. All cross
sections are reparametrized as a function of the invariant energy
above threshold $\sqrt{s}-\sqrt{s_0}$ \ \cite{cass99}, where
$\sqrt{s_0}$ denotes the threshold for the individual channel
given by the sum of the hadron masses in the final state of the
reaction. We recall that the differential cross sections for
binary reactions are fully determined by the total cross sections
when assuming $S$-wave dominance.

The question arises, furthermore, about the in-medium production
cross sections - essentially at density $\rho_0$ - if potentials
or self energies are involved. Here we employ the assumption that
the production matrix element squared $|M|^2$ does not change in
the medium and the change of cross section can be described by a
change of the available phase space. This notion is guided by the
experimental observation that meson production from $pp$
collisions is well described by phase space \cite{Moskal} if final
state interactions (FSI) between the hadrons are taken into
account. These FSI, furthermore, are found to be dominated by the
final baryon-baryon interaction, which is very strong in the $pp,
pn$ or $p \Lambda$ channels in free space. On the other hand, such
FSI's are screened in the nuclear medium due to the surrounding
nucleons such that in-medium production cross sections for mesons
are expected to vary essentially with the available phase space.
It is thus sufficient to shift the threshold in a $pN$ collision
to
\begin{equation}
\sqrt{s_0^*} = \Pi^0_N(p_N) + \Pi^0_\Lambda (p_\Lambda) + \Pi^0_K
(p_K) \label{shift}
\end{equation}
using (\ref{g26}), where the momenta $p_j$ denote
the relative momenta with respect to the nuclear matter rest frame
and $\Pi^0_j(p)$ is the energy of hadron $j$ including the
potentials (4).

\section{Comparison to experimental spectra}
In this Section we show the influence of nucleon and kaon
potentials on $K^+$ spectra from $p+^{12}C$ and $p+^{197}Au$
reactions and compare our calculations to the experimental $K^+$
spectra available from 1.0 GeV $-$ 2.3 GeV bombarding energy  from
the ANKE collaboration \cite{Markus04} that has taken  $K^+$
spectra in forward direction for $\theta_{lab} \leq 12^o$ up to
momenta of $\sim$ 500 MeV/c.

\begin{figure}[h]
\centerline{\psfig{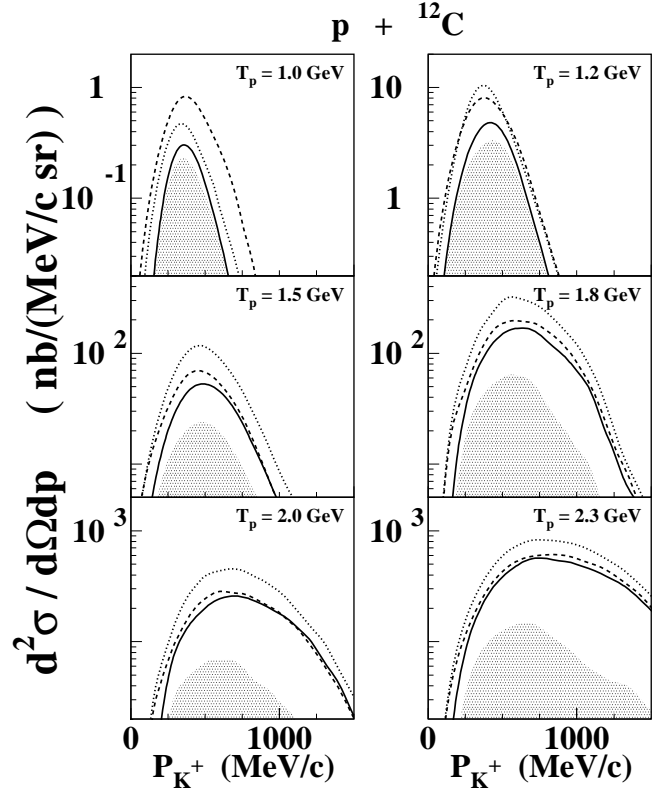}}
 \caption{The
calculated differential $K^+$ spectra for $p+^{12}C$ from 1.0 to
2.3 GeV for $\theta_{lab} \leq 12^0$. The dotted line is obtained
from CBUU calculations without baryon and kaon potentials, the
dashed line shows the results with baryon potentials included
while the solid line corresponds to calculations including both,
nucleon and kaon potentials (+20  MeV at nuclear density
$\rho_0$). The shaded areas indicates the contributions from the
two-step mechanisms $pN \rightarrow \Delta N, \Delta N \rightarrow K^+ Y N$
and $pN \rightarrow  \pi NN, \pi N
\rightarrow K^+ Y$, respectively, for the case of nucleon and kaon
potentials, such that the difference to the solid line corresponds
to the primary $pN$ production channel. }
 \label{bild1b}
\end{figure}

\begin{figure}[h]
\centerline{\psfig{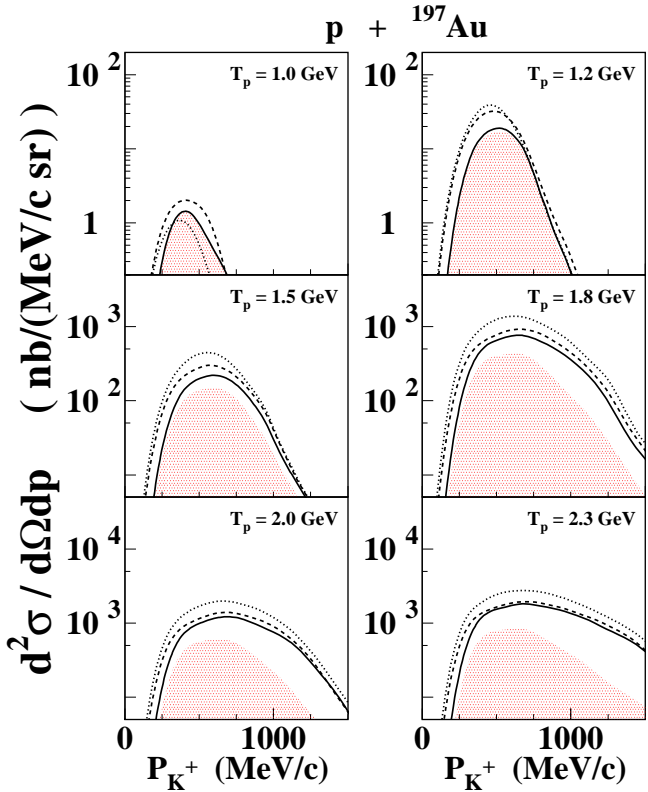}}
 \caption{The same as Fig. 1 for $p+Au$ reactions from 1.0 to
 2.3 GeV bombarding energy.}
 \label{bild1c}
\end{figure}
\begin{figure}[h]
\centerline{\psfig{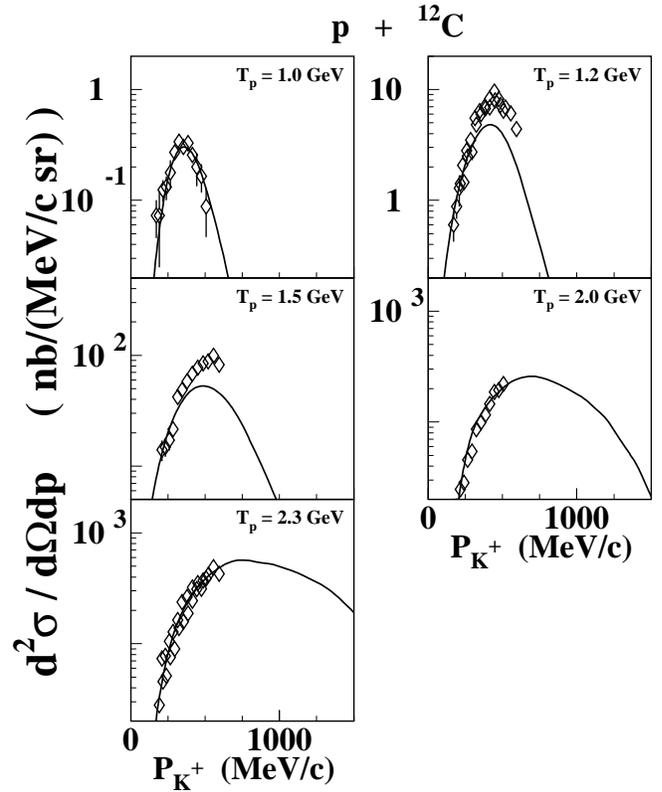}}
 \caption{Comparison of the $K^+$ spectra from the CBUU calculations (solid lines)
 with the data from the ANKE collaboration \cite{ANKE,Markus04}
 for $p+^{12}C$. The calculations include both
 the momentum-dependent nucleon potential as well as a repulsive
 $K^+$ potential of +20 MeV at density $\rho_0$.}
 \label{bild1d}
\end{figure}

The calculated differential $K^+$ spectra for $p+^{12}C$ from 1.0
GeV to 2.3 GeV for $\theta_{lab} \leq 12^0$  are displayed in Fig.
\ref{bild1b} and for $p+^{197}Au$ in Fig. \ref{bild1c}. The dotted
lines are obtained from CBUU calculations without baryon and kaon
potentials, the dashed lines show the results with baryon
potentials included while the solid lines correspond to
calculations with both, nucleon and kaon potentials. In the latter
case we have assumed a kaon potential of +20 MeV at density
$\rho_0$. At the low bombarding energy of 1.0 GeV the nuclear
potential in the final state is dominantly attractive and we
obtain an enhancement of  the $K^+$ yield by about a factor of 2
-- relative to the calculations without nuclear potential --
whereas the additional repulsive $K^+$ potential leads again to a
decrease by a factor $\sim$ 2-3. For 1.2 GeV the net effect of the
nucleon potential is close to zero; its repulsion in the final
state for higher bombarding energies becomes visible in a
reduction of the $K^+$ spectrum relative to the 'free' case. On
the other hand the effect of the repulsive kaon potential is
always a reduction of the $K^+$ spectra, which is most pronounced
at very low bombarding energies -- when comparing the solid lines
relative to the dashed lines -- and becomes small at high
bombarding energies.

The shaded areas in Figs. \ref{bild1b} and \ref{bild1c} indicate
the contributions from the two-step mechanisms
$pN \rightarrow \Delta N, \Delta N \rightarrow K^+ Y N$
and $pN \rightarrow  \pi NN, \pi N
\rightarrow K^+ Y$,  respectively,
for the case of nucleon and kaon potentials included (solid line).
Thus the role of secondary ($\Delta$ and pion induced) reaction
channels is clearly visible from Figs. 1 and 2 by comparing the
shaded area to the solid line, that correspond to the total
spectra. At $T_{lab}$ = 1.0 GeV the dominant fraction of the $K^+$
yield is due to the secondary channels in line with the earlier
calculations in Refs. \cite{15,cass95}. Consequently, one does not
probe high momentum components of the nuclear wave function by
$K^+$ spectra in a direct way. However, with increasing bombarding
energy the role of the two-step reactions diminishes gradually. At
$T_{lab}$ = 2.3 GeV the secondary channels in case of a $Au$
target amount to about 30\% and in case of a $C$ target to less
than 20\%. This relative change with target mass number is
attributed to the fact that for the small $^{12}C$ target only a
fraction of the high energy pions rescatters in the target and
produces $K^+ Y$ pairs. Moreover, the role of the secondary
channels decreases with increasing kaon momentum such that the
high momentum  tail of the $K^+$ spectra is dominated by the first
chance $pN$ production channel in line with the studies by Paryev
\cite{paryev}.

How do these calculations compare to the data from the ANKE
collaboration \cite{ANKE,Markus04} and what can we learn from
that? To answer this question we show in Fig. \ref{bild1d} our
calculations that include the baryon and $K^+$ potentials (solid
lines) with the measured spectra for $p+^{12}C$ from Refs.
\cite{ANKE,Markus04}. Whereas the description of the experimental
spectrum looks almost perfect at 1.0 GeV the data are slightly
underestimated at $T_{lab}$ = 1.2 and 1.5 GeV in the maximum.
At these energies the maximum of the spectrum is better described
without kaon and nuclear potentials, but then
the spectra no longer match
in the low momentum part. Thus a repulsive $K^+$ potential
is needed to properly describe the shape of the spectrum
especially for low momenta. On the other hand, channels such as
$pN_1 \rightarrow d+\pi$ followed by a secondary reaction $\pi
+N_2 \rightarrow K^+ + Y$ are not included in the present
transport calculations. Such channels have been found in Ref.
\cite{Vladimir} to contribute to the kaon spectrum at bombarding
energies $\sim$ 1.2 GeV but to be insignificant at higher energies
of 2.0 to 2.3 GeV. As can be seen from Fig. \ref{bild1d} the
calculations -- including nucleon and kaon potentials -- are in a
good agreement with the experimental spectra again at 2.0 and 2.3
GeV which is in line with the findings of Ref. \cite{Vladimir}.

\begin{figure}[h]
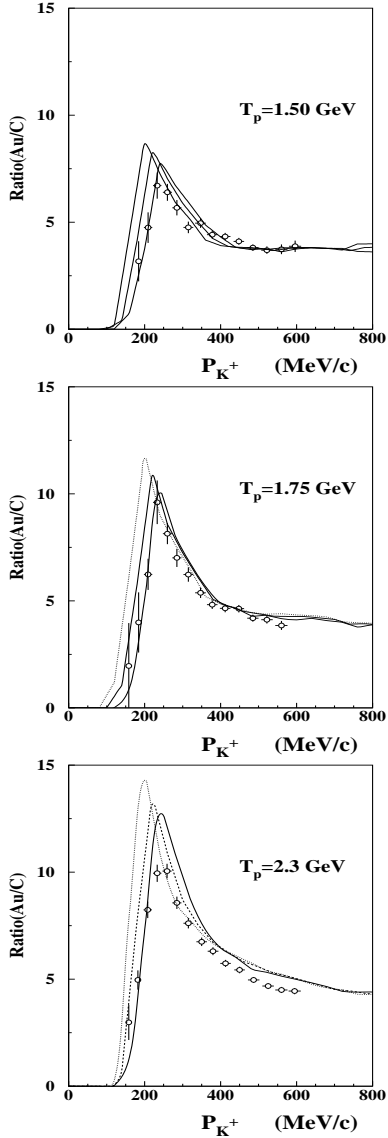

\centerline{\psfig{figure=fig3.epsi,width=5.0cm,height=5cm}}
\centerline{\psfig{figure=fig2.epsi,width=5.0cm,height=5cm}}
\centerline{\psfig{figure=fig1.epsi,width=5.0cm,height=5cm}}
 \caption{The ratio of the
calculated differential $K^+$ spectra from $p+Au$ to $p+C$
reactions at 1.5, 1.75 and 2.3 GeV in comparison to the data from
Ref. \cite{Markus04}. The different solid lines correspond to
calculations employing different kaon potentials at normal nuclear
density $\rho_0$. Starting from the left the results correspond to
repulsive kaon potentials of 0, 10 and 20 MeV at $\rho_0$.}
 \label{bild5}
\end{figure}

\section{$K^+$ spectral ratios}
The differential $K^+$ momentum spectra in Figs. 1 to 3 differ in
their low momentum part -- at fixed bombarding energy -- when
varying the target mass number. In Refs. \cite{zibi02,Mischa} it
has been pointed out that the variation in the low $K^+$ momentum
spectrum is a consequence of i) the different acceleration of the
charged kaon in the Coulomb field of the target and ii) an
additional acceleration in the repulsive kaon potential in the
nuclear medium (target nucleus). The combined effect of both the
Coulomb and nuclear repulsion is most easily seen when comparing
the $K^+$ spectra from different targets. To this aim we show in
Fig. \ref{bild5} the ratio of the spectra from $Au$ to $C$ targets
at 1.5, 1.75 and 2.3 GeV, where this effect is most pronounced
(cf. Ref. \cite{Mischa}). The different solid lines in Fig.
\ref{bild5} correspond to calculations employing different nuclear
kaon potentials  of magnitude 0, 10 and 20
MeV at normal nuclear density $\rho_0$ (starting from the left).

It is seen that the ratio for (Au/C) is approximately independent
on the bombarding energy and shows a maximum at about 230 MeV/c.
When neglecting kaon repulsion from nuclear forces
-- including only Coulomb repulsion -- this maximum shows up
slightly below 200 MeV/c, which is clearly too low in comparison
to the data (cf. \cite{Markus04,Mischa}). We recall that the
relative strength of the momentum shift in the forward $K^+$
spectra is proportional to the square root of the sum of Coulomb
and nuclear potentials, \begin{equation} \label{shift2} \Delta p
\approx \sqrt{2 M_K (U_{Coul}+U_K)}, \end{equation} which implies
that an additional repulsive nuclear $K^+$ potential is needed to
describe the data. Since the strength of this potential -- linear
in the nuclear density at the $K^+$ production point -- is a
'parameter' in the transport calculations, we may in turn
determine the strength of this potential in comparison to the
data. A $\chi^2$-fit gives $U_K(\rho_0) \approx$ 20 $\pm$ 5 MeV
for the ratio of $p+Au$ to $p+C$ reactions.

We mention that the transport calculations have a limited accuracy
due to low statistics in some regions of phase space. This holds
true especially for the low momentum spectrum of the kaons in the
acceptance of the ANKE spectrometer and has an impact also on the
ratio of the spectra from different targets. In the $\chi^2$-fit
addressed above  the statistical inaccuracy of the
calculations has been included.

\begin{figure}[h]
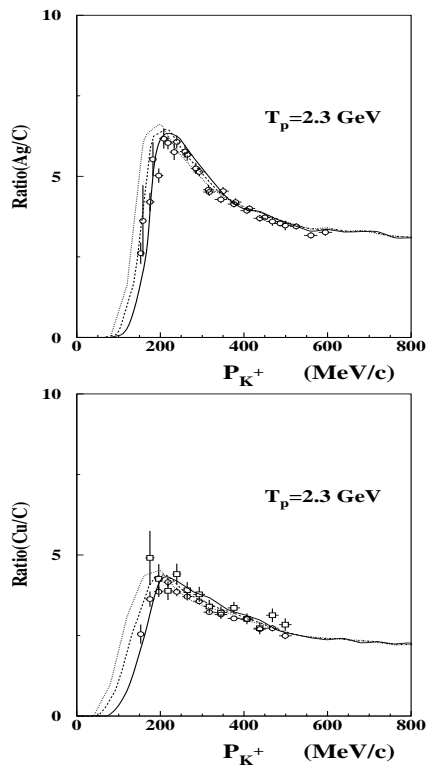

\centerline{\psfig{figure=fig4.epsi,width=5.5cm,height=5cm}}
\centerline{\psfig{figure=fig5.epsi,width=5.5cm,height=5cm}}
 \caption{The same ratios as in Fig. \ref{bild5} but for $Ag/C$ and $Cu/C$
 at 2.3 GeV in comparison to the data from Ref. \cite{Markus04}.}
 \label{bild6}
\end{figure}

It is of further interest whether the interpretation of the momentum
differential kaon spectrum -- discussed above -- is compatible
with the ratios from other targets, too. To this aim we show in
Fig. \ref{bild6} the same ratios as in Fig. \ref{bild5} but for
$Ag/C$ and $Cu/C$
 at 2.3 GeV in comparison to the data from Ref. \cite{Markus04}. It
 is clearly seen that the experimental ratios are consistent with
 the transport calculations for $U_K(\rho_0)$ = +20 MeV though the
 sensitivity to the strength of the kaon potential is reduced.
 This is basically due to the fact that the surface area
 --  relative to the total volume -- of the
 lighter targets is larger than for the heavy $Au$ nucleus.

 Nevertheless, the agreement between the CBUU calculations and the
 data is remarkable and demonstrates that the underlying $K^+$
 dynamics is rather well understood.

\section{Summary}
In this work we have studied the production of $K^+$  mesons in
proton-nucleus collisions from 1.0 to 2.3 GeV   with respect to
one-step nucleon-nucleon and two-step $\Delta$-nucleon or
pion-nucleon production channels on the basis of a coupled-channel
transport approach (CBUU) including differential transition
probabilities from $\pi N$ reactions (cf. Ref. \cite{zibi02}). We
have included the kaon final state interactions, which are
important for heavy targets, and incorporated the effects of
momentum-dependent potentials for the nucleon, hyperon and kaon in
the nucleus. A comparison of the calculations to the experimental
$K^+$ spectra taken at  COSY-J\"ulich has shown that the momentum
differential $K^+$ yield can be reproduced in magnitude and shape
within an accuracy better than 30\% at all bombarding energies.

The detailed calculations  have demonstrated that the spectra show
a substantial sensitivity to the potentials and their momentum
dependence. At low bombarding energies of $\sim$ 1.0 GeV the net
attractive potentials for the nucleon and the $\Lambda$-hyperon in
the final state lead to a relative enhancement of the $K^+$
spectra while at higher bombarding energies ($\sim$ 2 GeV) the
baryon potentials are repulsive and thus suppress $K^+$ production
relative to the free case. This phenomenon is seen in the
excitation function of the $K^+$ cross section when varying
$T_{lab}$ from 1.0 -- 2.3 GeV (cf. Fig. 3). Furthermore, the shape
of the spectrum for low $K^+$ momenta in the laboratory is very
sensitive to both, Coulomb and nuclear kaon potentials, since the
kaons are accelerated by both forces when leaving the nuclear
environment and propagating to the continuum. The relative
strength of this momentum shift in the forward $K^+$ spectra is
proportional to the square root of the sum of both potentials
(\ref{shift2}). Thus the $K^+$ spectral shape at low momenta (or
kinetic energies $T_K$) allows to determine the strength of the
$K^+$ potential from experimental data in an almost model
independent way especially when comparing kaon spectra from light
and heavy targets at the same bombarding energy (cf. Refs.
\cite{Markus04,Mischa,Barsov2}). In fact, the shape of such ratios
is practically independent of bombarding energy but sensitive to
the strength of the kaon repulsion in the medium. Our systematic
study has given a value of 20 $\pm$ 5 MeV at normal nuclear
density $\rho_0$ which agrees with the result from the
scattering length approximation \cite{Lutz}.

We  point out that our calculations slightly underpredict the
spectra at 1.2 and 1.5 GeV indicating that additional reaction
channels should be at work which have not been included in the
CBUU calculations. In fact, the experimental studies by the ANKE
collaboration in Ref. \cite{Vladimir} have shown that there is an
additional $K^+$ production from the secondary channel $p+N_1
\rightarrow d+\pi$; $\pi +N_2 \rightarrow K^+ + \Lambda$ at a
bombarding energy of 1.2 GeV in $p+C$ reactions. However, no
signals for this channel have been seen at higher bombarding
energies of $\sim$ 2 GeV. This interpretation is well in line with
our present transport calculations since the description of the
available spectra at 2.0 and 2.3 GeV for $p+^{12}C$ implies that
further reaction channels do not contribute significantly at the
higher bombarding energy. It remains presently an open question
whether a cooperative production mechanism by two- or three-nucleon
clusters --  not included in our transport calculations -- might
contribute and be identified from the experimental side in $p+A$
reactions at very low energies.

\vspace{0.5cm} The authors like to acknowledge valuable
discussions with  M. B\"uscher, V. Koptev, M. Nekipelov,
 P. Senger, A. Sibirtsev, H. Str\"oher and C. Wilkin on
various issues of this study.

\end{document}